\documentclass[prd,aps,showpacs,secnumarabic,superscriptaddress]{revtex4}
\usepackage{amsmath}
\usepackage[dvips]{graphicx,color}
\def\al{\alpha}

\def\ka{\kappa}

\def\ds{\displaystyle}

\def\ww{\mbox{\rule{18pt}{0pt}}}
\makeatletter
  \newcommand\figcaption{\def\@captype{figure}\caption}
  \newcommand\tabcaption{\def\@captype{table}\caption}
\makeatother
\begin{document}
\title{Critical Charges on Strange Quark Nuggets and Other Extended Objects}

\author{Duane A. Dicus}\email{dicus@physics.utexas.edu}
\affiliation{Physics Department, University of Texas, Austin, TX 78712}
\author{Wayne~W.~Repko}\email{repko@pa.msu.edu}
\affiliation{Department of Physics and Astronomy, Michigan State University, East Lansing, Michigan 48824, USA}
\author{V. L. Teplitz}\email{teplitz@milkyway.gsfc.nasa.gov}
\affiliation{NASA Goddard Space Flight Center, Greenbelt, MD 20771}
\affiliation{Physics Department, Southern Methodist University, Dallas, TX 75275}
\date{\today}
\begin{abstract}
We investigate the behavior of the critical charge for spontaneous pair production, $Z_C$, defined as the charge at which the total energy of a $K$-shell electron is $E=-m_e$, as a function of the radius $R$ of the charge distribution. Our approach is to solve the Dirac equation for a potential $V(r)$ consisting of a spherically symmetrical charge distribution of radius $R$ and a Coulomb tail. For a spherical shell distribution of the type usually associated with color-flavor locked strange quark nuggets, we confirm the relation $Z_C=0.71\,R({\rm fm})$ for sufficiently large $R$ obtained by Madsen, who used an approach based on the Thomas-Fermi model. We also present results for a uniformly charged sphere and again find that $Z_C\sim R$ for large enough $R$. Also discussed is the behavior of $Z_C$ when simple {\it ad hoc} modifications are made to the potential for $0\leq r < R$.
\end{abstract}
\maketitle

\section{Introduction}

In a recent paper \cite{SQN}, we examined the changes that develop on strange quark nuggets (SQN's) in space environments and discussed various signatures based on the charges they would acquire in these circumstances. Among these was the possibility that the surface charge on the nugget might exceed the critical charge $Z_C$ associated with the onset of spontaneous pair production. The condition for this to occur, that the energy of a $K$-shell electron satisfies $E=-m_e$, depends on the radius of the charge distribution \cite{PG,P,ZP}. For color-flavor locked SQN's \cite{RW}, it has been shown that the charge and the baryon number are related as $Z_Q=0.3\,A^{2/3}\sim\,R^2$ \cite{Mad1} and the question arises as to whether the nugget charge $Z_N$ ever exceeds the critical charge for any value of $R$. This issue was examined in a paper by Madsen \cite{Mad2}, who solved the Thomas-Fermi model for the case of the critical energy $E=-m_e$. He showed that the color-flavor locked relation $Z_C=Z_Q=0.3\,A^{2/3}$ is valid for nugget radii up to about $40\;{\rm fm}$ at which point there is a transition to a relation between the critical charge $Z_C$ and the radius $R$ given by $Z_C\al=2m_eR$ or $Z_C= 0.71\,R({\rm fm})$ for $R>400\,{\rm fm}$.

Our approach to the determination of the critical charge for a spherically symmetrical distribution is one of solving the Dirac equation with the energy fixed at the critical value $E=-m_e$. By matching the ratios of the radial solutions for $r<R$ and $R>r$ at the boundary, it is then possible to determine $Z_C$ for a given $R$ to any desired accuracy. In this way, the variation of $Z_C$ as a function of $R$ can be obtained. For the spherical shell, we find that the relation $Z_C\al=2m_eR$ is satisfied for $R\gtrsim 800\,{\rm fm}$. For $R\lesssim 200\,{\rm fm}$, $Z_C$ is slowly varying with respect to $R$. In Section \ref{sphere}, we derive the dependence of $Z_C$ on $R$ for two spherically symmetrical distributions, a charged shell and  a uniformly charged sphere. This is followed, in Section \ref{discuss}, by a discussion and some conclusions.

\section{Evaluating the Critical Charge for Spherical Distributions \label{sphere}}
In the following cases, we treat potentials of the form
\begin{equation}\label{V(r)}
V(r)=\left\{\begin{array}{ccc}
     {\cal V}(r)             & {\rm for}  & 0\leq r < R \\ [4pt]
    - \frac{\ds Z\al}{\ds r} & {\rm for}  & r> R
     \end{array} \right.\,,
\end{equation}
where ${\cal V}(r)$ takes on different forms depending on the particular charge distribution. To determine the critical charge, we solve the Dirac equation, whose upper and lower components $G(r)$ and $F(r)$ satisfy \cite{BjD}
\begin{subequations}\label{GF}
\begin{eqnarray}
\frac{dG(r)}{dr}&=&-\frac{\kappa}{r}G(r)+(m+E-V(r))F(r)      \label{G1} \\
\frac{dF(r)}{dr}&=& \frac{\kappa}{r}F(r)+(m-E+V(r))G(r) \,,  \label{F1}
\end{eqnarray}
\end{subequations}
in the special case $E=-m$. Here $\ka=\ell$ for $j=\ell-\frac{1}{2}$ and $\ka=-(\ell+1)$ for $j=(\ell+\frac{1}{2})$.

Depending on ${\cal V}(r)$, the form of the interior solution $(0\leq r <R)$ varies and, in the case of the uniformly charged sphere, must be obtained numerically. The form of the exterior $(r>R)$ solution is known \cite{P}. By setting $E=-m$ in Eqs.\,(\ref{GF}) and eliminating $F(r)$ in favor of $G(r)$, one obtains the differential equation
\begin{equation}\label{Gex}
x^2\frac{d^2}{dx^2}G(x)+x\frac{d}{dx}G(x)-x^2G(x)+4(Z^2\alpha^2-\kappa^2)G(x)\,=\,0\,,
\end{equation}
with $x=\sqrt{8Z\alpha\,r\,m}$. The solution to Eq.\,(\ref{Gex}) is a modified Bessel function of imaginary order
\begin{equation}\label{Gex(x)}
G(x)=K_{i\nu}(x)\quad {\rm with}\quad \nu=2\sqrt{Z^2\al^2-\ka^2}\,.
\end{equation}

To determine the critical charge, we use the condition that the ratio $F(r)/G(r)$ for the exterior and interior solutions must be equal at the boundary $r=R$ \cite{Rose}. For the exterior solution, this ratio can be obtained using Eq.\,(\ref{Gex(x)}) and Eq.\,(\ref{G1}) with $E=-m$. The result is
\begin{equation} \label{F/G_ex}
\frac{\ds F(r)}{G(r)}=\frac{\ka}{Z\al}+\frac{\ds r}{\ds Z\al\,G(r)}\frac{dG(r)}{dr}=\frac{\ka}{Z\al}+\frac{\ds\sqrt{8Z\al mr}}{2Z\al} \frac{\ds K'_{i\nu}(\sqrt{8Z\al mr})}{\ds K_{i\nu}(\sqrt{8Z\al mr})}\,.
\end{equation}
\subsection{Spherical Shell \label{shell}}
\begin{figure}[h]
\begin{minipage}[t]{0.45\textwidth}
  \centering
  \includegraphics[width=3in]{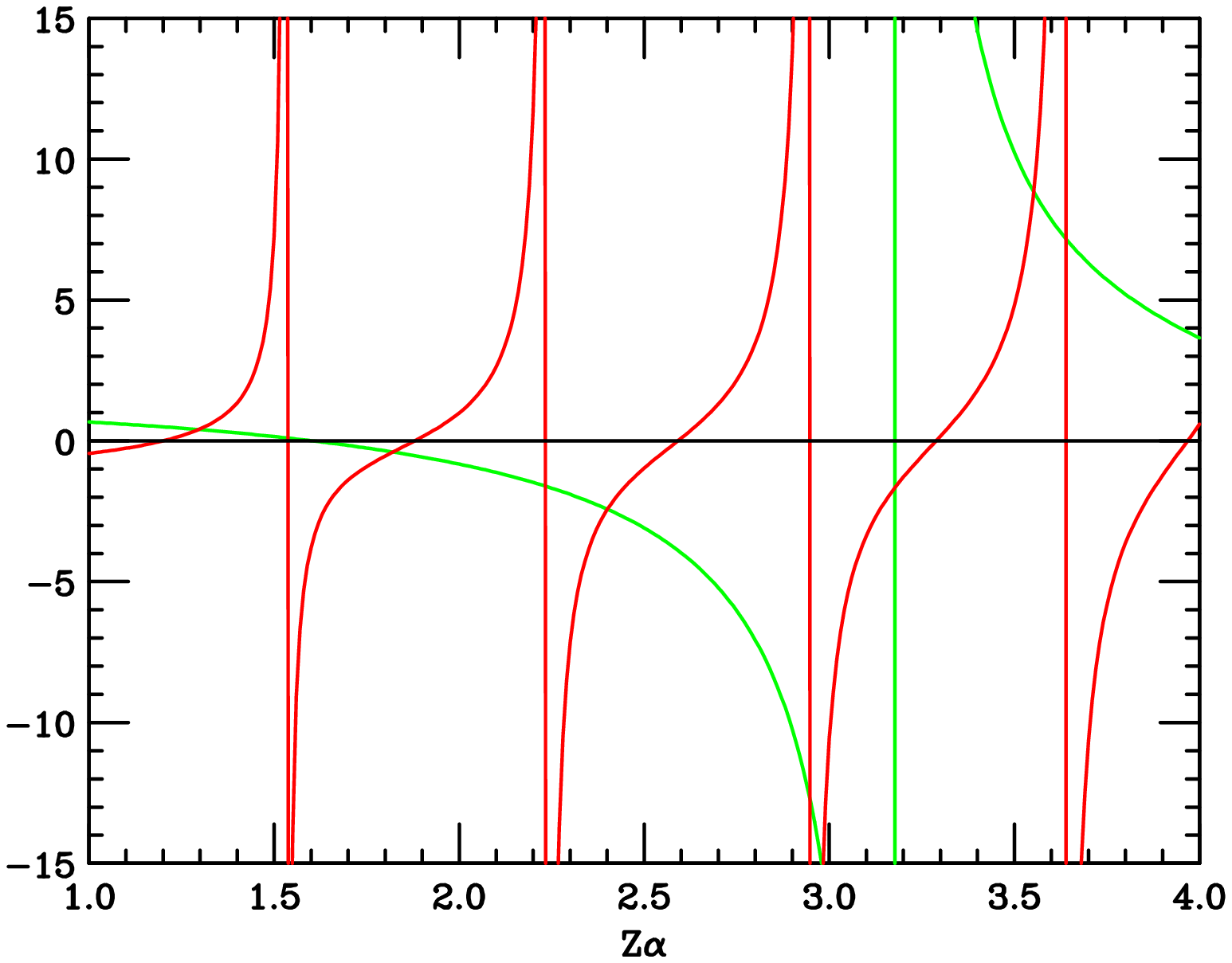}
  \caption{\footnotesize The solutions for the critical value of $Z\alpha$ corresponding to the $K$-shell is shown. The red curves are the exterior solution for $\hat{R}=0.25$ and the green curves are the corresponding interior solution.  \label{crit_shell}}
\end{minipage}%
\begin{minipage}[t]{0.1\textwidth}
\hfil
\end{minipage}%
\begin{minipage}[t]{0.45\textwidth}
  \centering
  \includegraphics[width=3in]{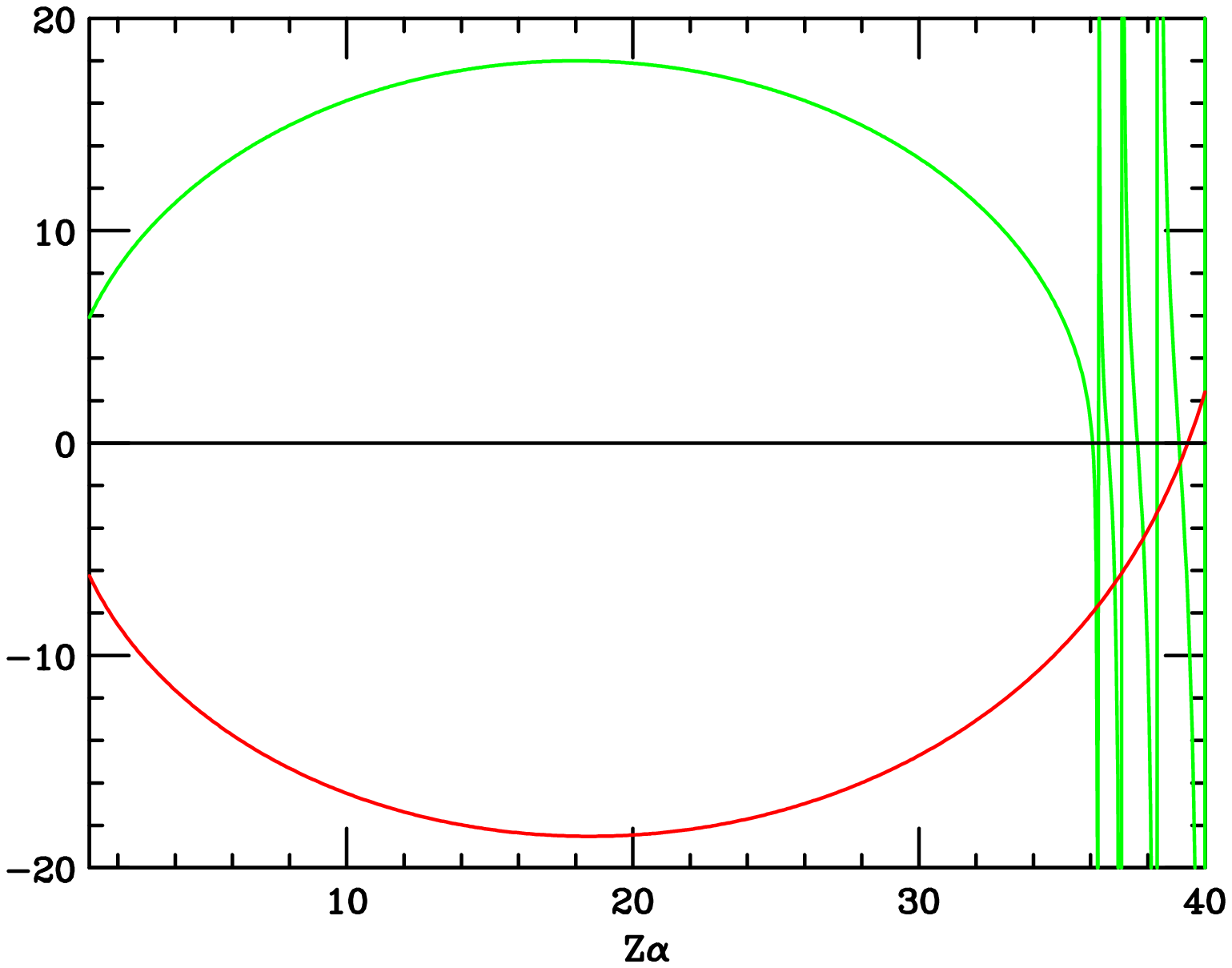}
  \caption{\footnotesize The same as Fig.\ref{crit_shell} for $\hat{R}=6.0$.  \label{crit_shell_6}}
\end{minipage}
\end{figure}

\begin{figure}[h]
\begin{minipage}[b]{0.4\textwidth}
\begin{tabular}{|r|r|r|r|}
\hline
\multicolumn{1}{|c|}{}        & \multicolumn{1}{c|}{$\ell=0$}& \multicolumn{1}{c|}{$\ell=1$} & \multicolumn{1}{c|}{$\ell=1$}  \\
\multicolumn{1}{|c|}{}        & \multicolumn{1}{c|}{$\ka=-1$}& \multicolumn{1}{c|}{$\ka=1$}  & \multicolumn{1}{c|}{$\ka=-2$}  \\
\hline
\multicolumn{1}{|c|}{$\hat{R}$} & \multicolumn{1}{c|}{$Z_C\al$}& \multicolumn{1}{c|}{$Z_C\al$}   & \multicolumn{1}{c|}{$Z_C\al$} \\
\hline
\ww 0.10\,\,  & \ww 1.154\,\,   & \ww 1.207\,\, & \ww 2.120\,\,   \\
0.25\,\,      & 1.296\,\,       & 1.414\,\,     & 2.251\,\,       \\
0.50\,\,      & 1.557\,\,       & 1.792\,\,     & 2.505\,\,       \\
1.00\,\,      & 2.275\,\,       & 2.730\,\,     & 3.206\,\,       \\
2.00\,\,      & 4.925\,\,       & 5.560\,\,     & 5.717\,\,       \\
3.00\,\,      & 9.608\,\,       & 10.147\,\,    & 10.191\,\,      \\
4.00\,\,      & 16.411\,\,      & 16.811\,\,    & 16.824\,\,      \\
5.00\,\,      & 25.291\,\,      & 25.586\,\,    & 25.590\,\,      \\
6.00\,\,      & 36.216\,\,      & 36.437\,\,    & 36.439\,\,      \\
7.00\,\,      & 49.166\,\,      & 49.337\,\,    & 49.336\,\,      \\
8.00\,\,      & 64.131\,\,      & 64.267\,\,    & 64.267\,\,      \\
9.00\,\,      & 81.106\,\,      & 81.216\,\,    & 81.216\,\,      \\
10.00\,\,     & 100.087\,\,     & 100.178\,\,   & 100.178\,\,     \\
\hline
\end{tabular}
\tabcaption{\footnotesize The dependence of $Z_C\al$ on $\hat{R}=\sqrt{2mR}$ is shown for the spherical shell distribution. The columns correspond to the $s_{1/2}$ $(\ka=-1)$, $p_{1/2}$ $(\ka=1)$ and $p_{3/2}$ $(\ka=-2)$ states. \label{shell_ZC}}
\par\vspace{0pt}
\end{minipage}%
\begin{minipage}[t]{0.1\textwidth}
\hfil
\end{minipage}%
\begin{minipage}[b]{0.5\textwidth}
  \centering
  \includegraphics[width=3in]{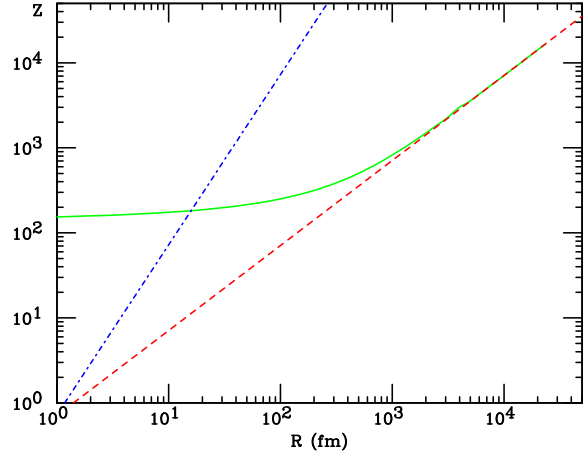} 
  \figcaption{\footnotesize The critical value of $Z$ for the spherical shell obtained by solving the Dirac equation (solid green line), the approach to Eq.\,(\ref{ZCvsR}) (red dashed line) and the expression for charge of the color-flavor locked SQN, $Z_Q=0.3\,A^{2/3}=0.726\,R^2$ (blue dot-dashed line) are shown. \label{ZCRZnug}}
\par\vspace{0pt}
\end{minipage}%
\end{figure}

For a constant interior potential, ${\cal V}(r)=V_0$, $G(r)$ satisfies
\begin{equation} \label{Gint}
G''(r)+\left(V_o(2m+V_o)-\frac{\kappa(\kappa+1)}{r^2}\right)G(r)\,=\,0\,.
\end{equation}
In particular, the interior potential for a uniformly charged spherical shell of radius $R$ is
\begin{equation} \label{Vshell}
{\cal V}(r)= -\frac{Z\al}{R}\,,
\end{equation}
and, taking $x=2mr$, Eq.\,(\ref{Gint}) becomes
\begin{equation} \label{Gin(x)}
G''(x)+\left(k^2-\frac{\ell(\ell+1)}{x^2}\right)G(x)\,.
\end{equation}
Here, we have used the fact that, for interior solutions which remain finite at $r=0$, $\ka(\ka+1)=\ell(\ell+1)$. The constant $k^2$ is
\begin{equation}
k^2=\frac{Z\al}{\hat{R}^2}\left(\frac{Z\al}{\hat{R}^2}-1\right)
\end{equation}
and $\hat{R}=\sqrt{2mR}$. The solution to Eq.\,(\ref{Gin(x)}) is $G(x)=xj_\ell(kx)$. Again, using Eq.\,(\ref{G1}), the ratio of the interior solutions is
\begin{equation} \label{F/G_in}
\frac{\ds F(r)}{G(r)}=\frac{\ka R}{Z\al r}+\frac{R}{Z\al r}\left((\ell+1)-2mkr\frac{\ds j_{\ell+1}(2mkr)}{\ds j_\ell(2mkr)}\right)\,.
\end{equation}
The critical value of $Z\al$ can now be determined by requiring that Eqs.\,(\ref{F/G_ex}) and (\ref{F/G_in}) are equal at $r=R$. The resulting condition is
\begin{equation}\label{shell_sol}
\ell+1-\sqrt{x(x-\hat{R}^2)}\,\frac{\ds j_{\ell+1}\left(\sqrt{x(x-\hat{R}^2)}\,\right)} {j_{\ell}\left(\sqrt{x(x-\hat{R}^2)}\,\right)} = \hat{R}\sqrt{x}\,\frac{\ds K'_{2i\sqrt{x^2-\ka^2}}(2\hat{R}\sqrt{x})}{K_{2i\sqrt{x^2-\ka^2}}(2\hat{R}\sqrt{x})}\,,
\end{equation}
where $x=Z\al$. Accurate numerical evaluations of the Bessel functions in Eq.\,(\ref{shell_sol}) are available in both the Mathematica and Maple packages. The location of the critical value of $Z\al$ can be estimated by plotting both sides of Eq.\,(\ref{shell_sol}) and then using a search routine to obtain the precise value. This is illustrated in Fig.\ref{crit_shell} for $R=12\,{\rm fm}$ $(\hat{R}=0.25)$ and $\ell=0$ ($\ka=-1$) which corresponds to the $s_{1/2}$ states. The lowest crossing, $Z\al=1.296$ in this case \cite{P}, determines the critical charge.
As $\hat{R}$ increases, the first solution continues to increase, approaching the value $Z\al=\hat{R}^2=2mR$. This is illustrated in Fig.\,\ref{crit_shell_6} for $\hat{R}=6$ and shown in detail in Table \ref{shell_ZC}.

If $R$ is expressed in femtometers, the relation between the critical charge $Z_C$ and $R$ can be written as
\begin{equation} \label{ZCvsR}
Z_C=0.71\,R\,(\mathrm{fm})\,,
\end{equation}
which is the result found by Madsen \cite{Mad2}, who derived it in a totally different way (Thomas-Fermi screening). He argues that Eq.\,(\ref{ZCvsR}) holds for $R>400\,\mathrm{fm}$ whereas we find that onset of validity is more like $R\gtrsim 800-1000\,\mathrm{fm}$, as can be seen in Fig.\,\ref{ZCRZnug}.
In this figure, we show the critical value of $Z$ obtained by solving the Dirac equation for $\ka=-1$ (solid green line) and the approach to Eq.\,(\ref{ZCvsR}) (red dashed line) together with the expression for charge of the color-flavor locked SQN, $Z_Q=0.3\,A^{2/3}=0.726\,R^2$ (blue dashed line).

For SNQ's with $R\gtrsim 20\,\mathrm{fm}$, the value of $Z$ obtained from the relation $Z_Q=0.3\,A^{2/3}$ exceeds the critical value and the nugget will screen its charge by producing pairs until its charge falls to the green curve in Fig.\,\ref{ZCRZnug}. As the nugget radius increases, the screening process continues to limit the nugget's charge to the values on the critical line. Ultimately, the relation between the critical charge and the nugget radius is given by Eq.\,(\ref{ZCvsR}). Beyond some radius, Madsen argues that the time required to screen the charge will be such that the relation between the charge and the radius will again be of the form $Z\sim\,R^2$. The analysis leading to this conclusion is quite unrelated to the task of determining the critical charge using the Dirac equation or the Thomas-Fermi model, and we have nothing to add in this regard beyond what is discussed in Ref.\cite{SQN}.

\subsection{Uniformly Charged Sphere \label{uniform}}

For a uniformly charged sphere of radius $R$, the interior potential is
\begin{equation} \label{Vsphere}
{\cal V}(r)=\frac{Z\al}{2R}\left(\frac{r^2}{R^2}-3\right).
\end{equation}
Inserting this potential into Eqs.\,(\ref{GF}) with $E=-m$ and letting $y=r/R$, the functions $G(y)$ and $F(y)$ satisfy
\begin{subequations}\label{GGFF}
\begin{eqnarray}
\frac{dG(y)}{dy}&=&-\frac{\kappa}{y}G(y)-\frac{x}{2}\left(y^2-3\right)F(y)  \label{GG1} \\
\frac{dF(y)}{dy}&=&\frac{\kappa}{y}F(y)+\left(\hat{R}^2+ \frac{x}{2}\left(y^2-3\right)\right)G(y)\,,  \label{FF1}
\end{eqnarray}
\end{subequations}
where, as before, $\hat{R}=\sqrt{2mR}$ and $x=Z\al$. In this case, the ratio of $F(r)/G(r)$ at the boundary $r=R$ $(y=1)$ must be obtained by numerically integrating Eqs.\,(\ref{GGFF}) for a given $R$ and a range of $Z\al$. This can be done using routines available in Mathematica or Maple and the result compared with the exterior ratio Eq.\,(\ref{F/G_ex}) to obtain an estimate of the lowest value of $Z\al$. The value can be refined to any desired accuracy by searching in the vicinity of the estimated solution.
\begin{figure}[h]
\begin{minipage}[b]{0.4\textwidth}
\begin{tabular}{|r|r|r|r|}
\hline
\multicolumn{1}{|c|}{}        & \multicolumn{1}{c|}{$\ell=0$}& \multicolumn{1}{c|}{$\ell=1$} & \multicolumn{1}{c|}{$\ell=1$}  \\
\multicolumn{1}{|c|}{}        & \multicolumn{1}{c|}{$\ka=-1$}& \multicolumn{1}{c|}{$\ka=1$}  & \multicolumn{1}{c|}{$\ka=-2$}  \\
\hline
\multicolumn{1}{|c|}{$\hat{R}$} & \multicolumn{1}{c|}{$Z_C\al$}& \multicolumn{1}{c|}{$Z_C\al$}   & \multicolumn{1}{c|}{$Z_C\al$} \\
\hline
\ww 0.10\,\,  & \ww 1.121\,\,   & \ww 1.190\,\, & \ww 2.111\,\,\\
0.25\,\,      & 1.267\,\,       & 1.372\,\,     & 2.228\,\,\\
0.50\,\,      & 1.487\,\,       & 1.692\,\,     & 2.445\,\,\\
1.00\,\,      & 2.059\,\,       & 2.460\,\,     & 3.019\,\,\\
2.00\,\,      & 4.095\,\,       & 4.651\,\,     & 4.929\,\,\\
3.00\,\,      & 7.241\,\,       & 7.982\,\,     & 8.124\,\,\\
4.00\,\,      & 11.87\,\,       & 10.67\,\,     & 12.71\,\,\\
5.00\,\,      & 17.86\,\,       & 16.67\,\,     & 18.67\,\,\\
6.00\,\,      & 25.18\,\,       & 24.00\,\,     & 25.98\,\,\\
7.00\,\,      & 33.83\,\,       & 32.67\,\,     & 34.36\,\,\\
8.00\,\,      & 43.83\,\,       & 42.67\,\,     & 44.63\,\,\\
9.00\,\,      & 55.17\,\,       & 54.00\,\,     & 55.95\,\,\\
10.00\,\,     & 67.83\,\,       & 66.67\,\,     & 68.61\,\,\\
12.00\,\,     & 97.16\,\,       & 96.00\,\,     & 97.94\,\,\\
14.00\,\,     & 131.83\,\,      & 130.67\,\,    & 132.60\,\,\\
16.00\,\,     & 171.83\,\,      & 170.67\,\,    & 172.60\,\,\\
18.00\,\,     & 217.16\,\,      & 216.00\,\,    & 221.02\,\,\\
20.00\,\,     & 267.82\,\,      & 266.67\,\,    & 268.60\,\,\\
\hline
\end{tabular}
\tabcaption{\footnotesize The dependence of $Z_C\al$ on $\hat{R}=\sqrt{2mR}$ is shown for the uniform charge distribution. The columns correspond to the  $s_{1/2}$ $(\ka=-1)$, $p_{1/2}$ $(\ka=1)$ and $p_{3/2}$ $(\ka=-2)$ states. \label{sphereZZC}}
\par\vspace{0pt}
\end{minipage}%
\begin{minipage}[t]{0.1\textwidth}
\hfil
\end{minipage}%
\begin{minipage}[b]{0.5\textwidth}
  \centering
  \includegraphics[width=3.25in]{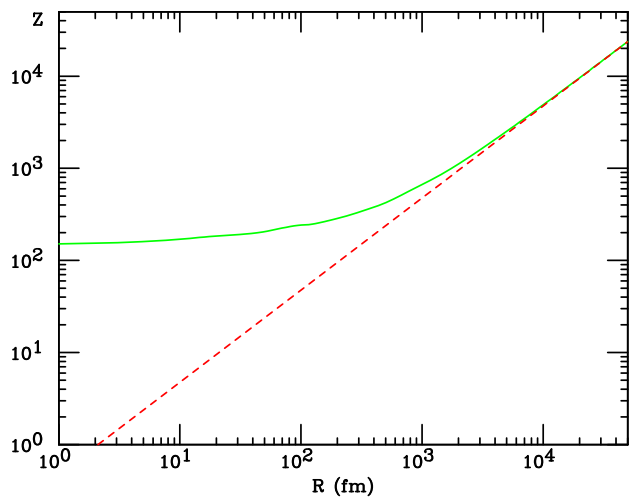}
  \caption{\footnotesize The critical $Z$ for the uniformly charged sphere of radius $R$ is shown in the solid green line. The dashed red line is $Z=0.475\,R$. \label{ZZCR}}
\par\vspace{0pt}
\end{minipage}
\end{figure}

Values of $Z_C\al$ for a range of $\hat{R}$ are given in Table \ref{sphereZZC} and the critical charge for $\ka=-1$ is shown as the solid green line in Fig.\,\ref{ZZCR}. The dependence of the critical charge on the radius is quite similar to the spherical shell case shown in Fig.\,\ref{ZCRZnug} in that there is a region of slow variation followed by a transition to a linear dependence on $R$. Overall, the values of $Z_C$ for the uniformly charge sphere are lower at a given $R$ than in the spherical shell case, and the asymptotic $R$-dependence is $Z_C\al\simeq \frac{4}{3}m_e R$.

\section{Discussion and Conclusion \label{discuss}}

We have obtained the critical charge as a function of the radius for two spherically symmetrical charge distributions by solving the Dirac equation to determine the values of $Z\al$ that produce $K$-shell and $L$-shell electron energies of $E=-m$. In both cases, the spherical shell and the uniformly charged sphere, the dependence of the critical charge on distribution's radius is similar.

\begin{figure}[h]
  \centering
  \includegraphics[width=3in]{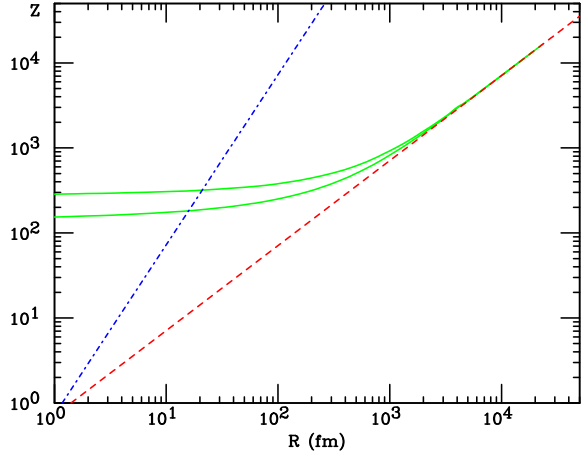}
  \caption{\footnotesize The behavior of the critical values of $Z
$ are shown as a function of the shell distribution radius $R$ for levels with $\ka=-1$ and $\ka=-2$ (solid green lines) as they approach the line $Z\al=2mR$ (red dashed line). The SQN charge, $Z_Q=0.3\,A^{2/3}$, is also shown (blue dot-dashed line). \label{ZSPR}}
\end{figure}

For the shell distribution, we exactly reproduce the large $R$ relation $Z_C\al=2mR$ found by Madsen \cite{Mad2}, who used an approach based on the Thomas-Fermi atomic model. We see this behavior for $s$-states as well as for $p$-states with $j=1/2$ and $3/2$. For radii of order a few fermis, the behavior of the critical value of $Z\al$ is controlled by the order of the modified Bessel function in Eq.\,(\ref{Gex(x)}), $\nu=2i\sqrt{Z^2\al^2-\ka^2}$. Unless $Z\al>|\ka|$, the Bessel function is not oscillatory and there is no solution to the boundary condition. Hence, the critical $Z\al$ begins at a value slightly larger than $|\ka|$ for small $R$, and evolves to the line $Z\al=2mR$ as $R$ increases. This behavior can be seen in Table\,\ref{shell_ZC} and is illustrated in Fig.\,(\ref{ZSPR}).

The situation for the uniformly charged sphere is quite similar. Generally speaking, the critical values of $Z\al$ for a given radius are slightly smaller than the corresponding values for the charged shell. For large $R$, the critical value of $Z\al$ is again proportional to $R$ with a smaller slope, $\frac{4}{3}m$. The details are given in Table \ref{sphereZZC}.

\begin{figure}[h]
\centering
\includegraphics[width=3in]{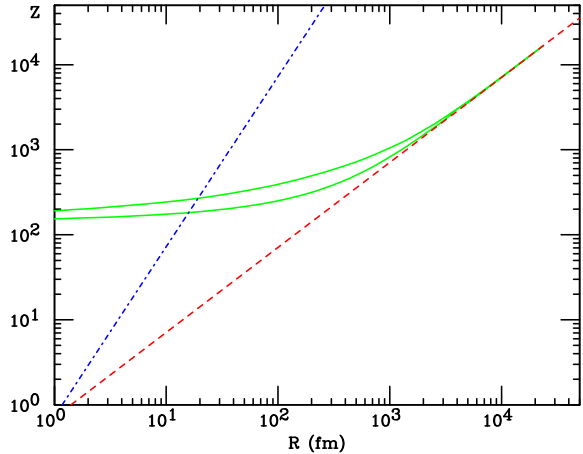}
\caption{\footnotesize The behavior of the first and second critical values of $Z$ for $\ka=-1$ are shown as a function of the spherical shell radius $R$  (solid green lines) as they approach the line $Z\al=2mR$ (red dashed line). The SQN charge, $Z_Q=0.3\,A^{2/3}$, is also shown (blue dot-dashed line). \label{1S-2S}}
\end{figure}

In the discussion thus far, we examined the $R$-dependence of the {\it lowest} critical value of $Z\al$ for a given value of $\ka$ and found a universal linear dependence for sufficiently large values of $R$. This behavior persists when higher critical values of $Z\al$ for a particular $\ka$ are examined, as illustrated in Fig.\ref{1S-2S}, where we plot the evolution of the first and 
second critical values of $Z$ for $\ka=-1$ as a function of $R$ in the spherical shell case.
 In fact, all states will have a $Z_C\al$ dependence on $R$ with the same general shape as the curves in Fig.\ref{1S-2S}; they will simply start at larger values on the vertical ($Z$) axis. We used Eq.\,(\ref{shell_sol}) at $R=2\,{\rm fm}$ with $\ka$ from $-1$ to $3$ ($\ell$ from 0 to 3) to do a count and found $64$ states with $Z_C\al<4$. These range from the $1s_{1/2}$ state with $Z_C=1.15$ to the $11f_{5/2}$ state with $Z_C\al=3.97$ and can accommodate 472 electrons.

We also looked at an {\it ad hoc} modification of the interior potential , \begin{equation}
{\cal V}(r)=-\frac{Z\al}{2R}\left(1+\frac{r^2}{R^2}\right),
\end{equation}
that is less negative at $r=0$ by a factor of two than the shell potential of Eq.\,(\ref{Vshell}), rather than being more negative as is the case for the uniform sphere potential of Eq.\,(\ref{Vsphere}). The resulting  values of $Z_C\al$ are slightly larger than those in Table \ref{shell_ZC} and have a very similar linear behavior for large enough $R$.

$Z_C$ is the value of $Z_N$ such that the least tightly bound electron in the ground state of the system has a binding energy of $-2m_e$. If that electron is removed, it becomes energetically favorable for the vacuum to produce an $e^+e^-$ pair. The interaction of the least bound electron with the other electrons is not taken into account in our calculations, but this should not change things by a great deal.

To summarize, solving the Dirac equation with a spherically symmetrical charge distribution of total charge $Ze$ and radius $R$ for the energy $E=-m$ predicts that the critical charge obeys $Z_C\sim R$ for sufficiently large $R$. This result holds for any value of the parameter $\ka$ as well as for any of the critical values of $Z\al$ associated with a given $\ka$. However, the approach to the linear dependence on $R$ and its slope both depend on the particular charge distribution.
\begin{acknowledgments}
DAD was supported in part by the U.S. Department of Energy under Grant No. DE-FG03-93ER40757.  WWR was supported in part by the National Science Foundation under Grant PHY-0555544.
\end{acknowledgments}

\end{document}